\documentclass[aps, prb, twocolumn, longbibliography]{revtex4-2}
\usepackage{amsmath,amssymb}
\usepackage[dvipdfmx]{graphicx}
\usepackage{bm}
\usepackage{braket}
\usepackage{color}

\newcounter{num}

\begin{document}
\title{Antisymmetric linear transverse magnetization and ferroaxial moments 
\\ induced by geometry-driven electric field gradients}

\author{Akane Inda and Satoru Hayami}
\affiliation{
Graduate School of Science, Hokkaido University, Sapporo 060-0810, Japan
}
\begin{abstract}
  We theoretically investigate the transverse magnetization and ferroaxial moments induced by electric field gradients arising from the geometry of finite systems.
  Based on the Kubo formalism and real-time numerical simulations for a finite trapezoidal model, we demonstrate that both quantities are generated under the electric field gradient and are enhanced by tuning the leg inclination, which controls the gradient strength.
  We further show that the induced transverse magnetization is antisymmetric and linear in the magnetic field; such a response is prohibited by Onsager reciprocity in the absence of an electric field gradient. 
  In addition, we find that the total transverse magnetization scales linearly with the electric field, in contrast to the longitudinal one, which exhibits a quadratic dependence, providing an advantage for experimental observation. 
  Our results establish geometry-induced electric field gradients as a versatile mechanism for realizing and controlling unconventional transverse responses in mesoscopic systems.
\end{abstract}

\maketitle
\section{Introduction}
\label{sec:intro}

In condensed matter physics, the responses of charge, spin, orbital, and lattice degrees of freedom to external fields are fundamentally governed by symmetry. 
The allowed response tensors are determined by the symmetry of the system, as well as by the symmetries and tensorial ranks of the external fields and observables~\cite{birss1964symmetry, litvin2013magnetic, gallego2019automatic, yatsushiro2021prb_122, etxebarria2025crystal}. 
While most theoretical studies have focused on bulk response to spatially uniform external fields in the long-wavelength limit, electromagnetic fields are often spatially nonuniform in realistic situations. 
Such spatial variations naturally arise near surfaces and interfaces, in mesoscopic systems, and under asymmetric electrode geometries. 
Importantly, spatially varying electromagnetic fields act as higher-rank perturbations, thereby enabling unconventional response phenomena, such as natural optical activity~\cite{landau2013electrodynamics}, gyrotropic birefringence~\cite{hornreich1968theory}, nonlinear nonreciprocal electric conductivity driven by magnetic field gradients~\cite{yamanaka2025nonlinear}, and strain gradient-induced magnetization~\cite{koyama2026flexocurrentmagnetization}.
Recent studies have revealed a quantitative correspondence between field-gradient responses and higher-rank multipole degrees of freedom~\cite{oike2025thermodynamic, shitade2025intrinsic, sato2026quantum}.
Despite these advances, the role of electric field gradients as a versatile control parameter remains largely unexplored.

\begin{figure}[t!]
  \centering
  \includegraphics[width=\linewidth]{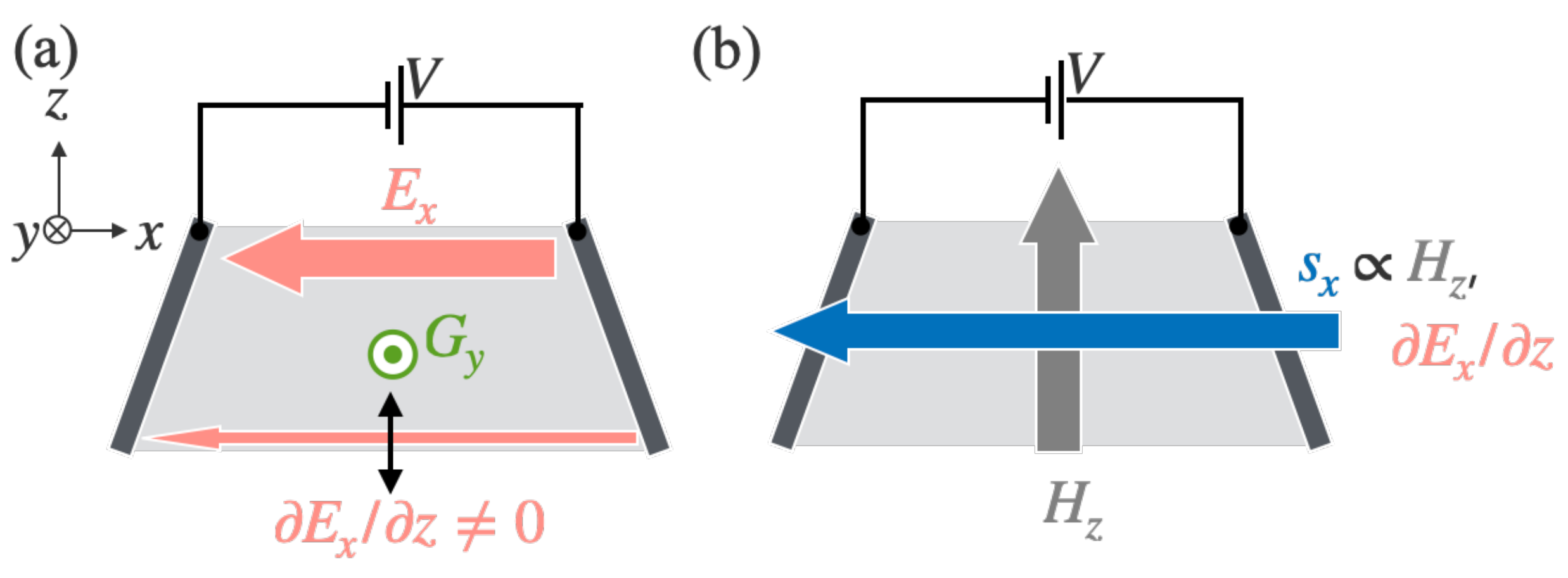}
  \caption{
  \label{f:intro}
  (a) Ferroaxial moment $G_y$ induced by the electric field gradient $\partial E_x/\partial z$ 
  (the component $\partial E_z/\partial x$ is omitted for visual clarity).
  (b) Transverse magnetization $s_x$ generated by the magnetic field $H_z$ under the electric field gradient.
  }
\end{figure}

In this paper, we theoretically investigate physical phenomena induced by electric field gradients that emerge from the geometry of finite systems. 
Considering the trapezoidal structure shown in Fig.~\ref{f:intro}(a), electric field gradients with perpendicular components, such as $\partial E_x/\partial z$ and $\partial E_z/\partial x$, naturally arise when a voltage is applied between the legs.
Such field configurations give rise to ferroaxial moments $\bm{G}$, which are time-reversal-even axial vectors~\cite{Hlinka2014prl_eight-types, hlinka2016aps_Symmetry_Guide_to_Ferroaxial_Transitions}, as shown in Fig.~\ref{f:intro}(a). 
Microscopically, the ferroaxial moment can be understood as a vortex structure of electric polarizations, $\bm{G} = \sum_i \bm{r}_i \times \bm{p}_i$, where $\bm{r}_i$ and $\bm{p}_i$ denote the position vector and electric polarization at site $i$, respectively~\cite{dubovik1990toroid}. 
This relation immediately indicates that $(\nabla \times \bm{E})_y$ acts as the conjugate field to $G_y$ from the symmetry aspect. 

Ferroaxial ordering has attracted growing interest owing to the emergence of intriguing physical properties, especially the transverse responses of the conjugate physical quantities~\cite{cheong2021permutable, Hayami2022jpsj_spincurrent}, including the spin current generation~\cite{Roy2022prm_spin-current, Hayami2022jpsj_spincurrent}, antisymmetric thermopolarization~\cite{nasu2022prb_thermopolarization}, nonlinear transverse magnetization (TM)~\cite{inda2023jpsj, du2026electric}, unconventional Hall effect~\cite{Hayami_PhysRevB.108.085124}, nonlinear magnetostriction~\cite{kirikoshi2023rotational}, and nonlinear Edelstein effect~\cite{kirikoshi2026light}. 
However, only a limited number of candidate materials have been identified compared with other ferroic orders, such as Cu$_3$Nb$_2$O$_8$~\cite{johnson2011prl_Cu3Nb2O8}, CaMn$_7$O$_{12}$~\cite{johnson2012prl_CaMn7O12}, BaCoSiO$_4$~\cite{xu2022prb_BaCoSiO4}, Ca$_5$Ir$_3$O$_{12}$~\cite{Hasegawa_doi:10.7566/JPSJ.89.054602, hanate2021first, hayami2023cluster, hanate2023space}, K$_2$Zr(PO$_4$)$_2$~\cite{yamagishi2023ferroaxial, Bhowal_PhysRevResearch.6.043141, xie2025spinless}, Na$_2$Hf(BO$_3$)$_2$~\cite{nagai2023chemicalSwitching}, Na-superionic conductors~\cite{nagai2023chemical}, MnTiO$_3$~\cite{Sekine_PhysRevMaterials.8.064406}, and 1$T$-TaS$_2$~\cite{Luo_PhysRevLett.127.126401, liu2023electrical}.
Moreover, experimental control remains challenging because ferroaxial moments do not directly couple to uniform electric or magnetic fields due to spatial inversion and time-reversal symmetries, making domain control and macroscopic detection difficult.

The present setup shown in Fig.~\ref{f:intro}(a) offers a way to overcome these difficulties, as the electric field gradient directly acts as the conjugate field of the ferroaxial moment.
An additional advantage is its generality: electric field gradients can be engineered geometrically in a wide range of insulating systems, largely independent of crystallographic symmetry.
This provides a versatile platform for inducing ferroaxial-related phenomena.
Although a time-dependent magnetic field $d\bm{B}/dt$ can also serve as a conjugate field via Maxwell's equations, generating sufficiently large magnetic-field frequencies is experimentally challenging, whereas electric field gradients are expected to be controllable~\cite{Prosandeev_PhysRevLett.96.237601,Ivan_PhysRevLett.101.197601}.

In addition to ferroaxial moments, we investigate the antisymmetric TM induced by the magnetic field under the electric field gradient, as shown in Fig.~\ref{f:intro}(b).
From the symmetry viewpoint, the antisymmetric component of the TM can, in principle, appear as a linear response to a magnetic field; however, such contributions are typically prohibited in the absence of the electric field gradient.
We demonstrate that the electric field gradient activates a TM that is linear in both the magnetic field and the electric field, while the longitudinal magnetization remains quadratic in the electric field. 
This distinction allows the transverse response to be selectively enhanced, providing a clear experimental advantage.

To elucidate these phenomena, we employ a tight-binding model on a finite trapezoidal system, as shown in Fig.~\ref{f:model}(a), and anlyaze it using both the Kubo formalism and real-time numerical simulations.
We show that the electric field gradient induces both magnetizations perpendicular to the magnetic field and ferroaxial moments. 
Furthermore, we clarify an antisymmetric TM that is linear in the magnetic field, which is forbidden by Onsager reciprocity in the absence of spatially nonuniform fields.
We also demonstrate that both ferroaxial moments and the TM can be enhanced by tuning the geometric parameter that controls the electric field gradient, namely the inclination of the trapezoidal legs.

The rest of this paper is organized as follows. 
In Sec.~\ref{sec:kubo}, we introduce the model Hamiltonian in a finite trapezoidal system. 
We then evaluate the susceptibility of the ferroaxial moment based on the Kubo formalism.  
In Sec.~\ref{sec:TM}, we perform real-time simulations to investigate the TM induced by the external magnetic field in the presence of the electric field gradient.
The dependence on the geometric parameter controlling the electric field gradient is discussed in Sec.~\ref{sec:slope_dependence}.
Finally, we summarize our results in Sec.~\ref{sec:summary}.
In Appendix~\ref{appendix:w_uniform_E}, we show the total ferroaxial moment and TM induced upon applying a voltage, including the contribution from the uniform electric field which is neglected in the main text.

\section{Ferroaxial moment induced by electric field gradient} 
\label{sec:kubo}
\begin{figure}[t!]
  \centering
  \includegraphics[width=\linewidth]{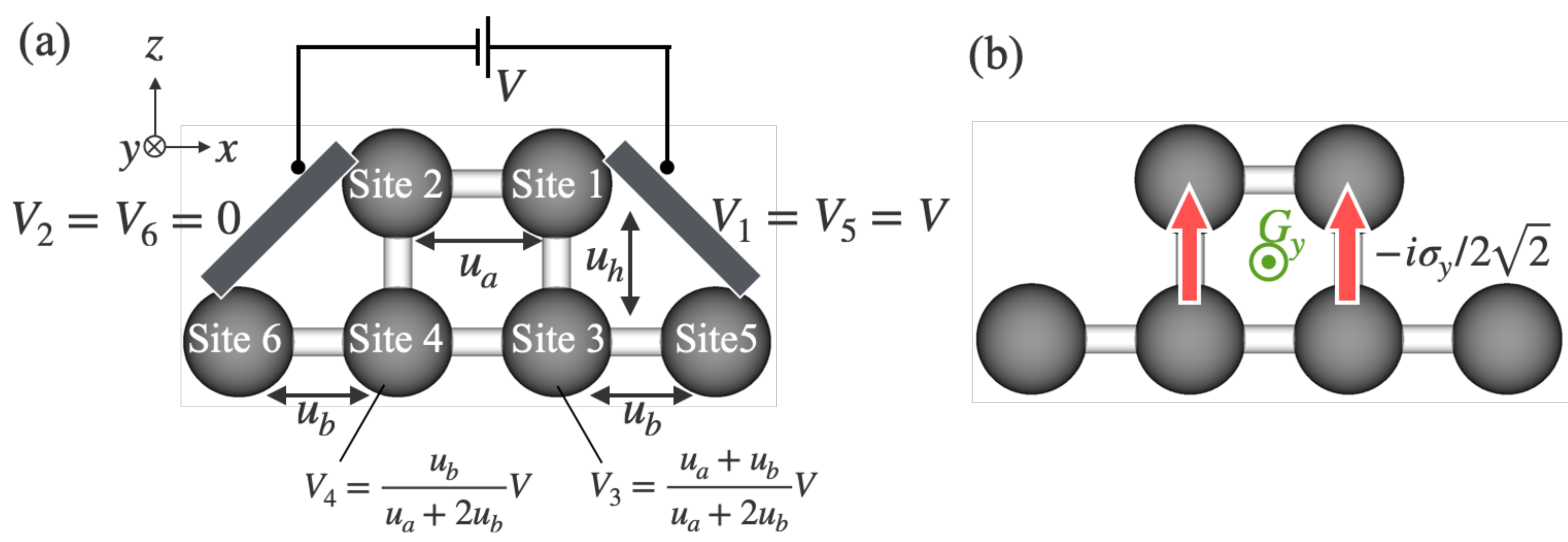}
  \caption{
  \label{f:model}
  Schematic pictures of (a) the trapezoidal system under the point group $C_{\rm 2v}$ and (b) the ferroaxial moment $G_y$. 
  The parameters $u_a$, $u_b$, and $u_h$ denote the characteristic lengths of the trapezoidal geometry and $V_i (i=1,2,3,4,5,6)$ is the electrostatic potential at site $i$.
  In (b), the out-of-plane ferroaxial moment $G_y$ (green circle) is represented by spin-dependent imaginary hoppings, as indicated by the pink arrows.
  }
\end{figure} 

\begin{figure}[t!]
  \centering
  \includegraphics[width=\linewidth]{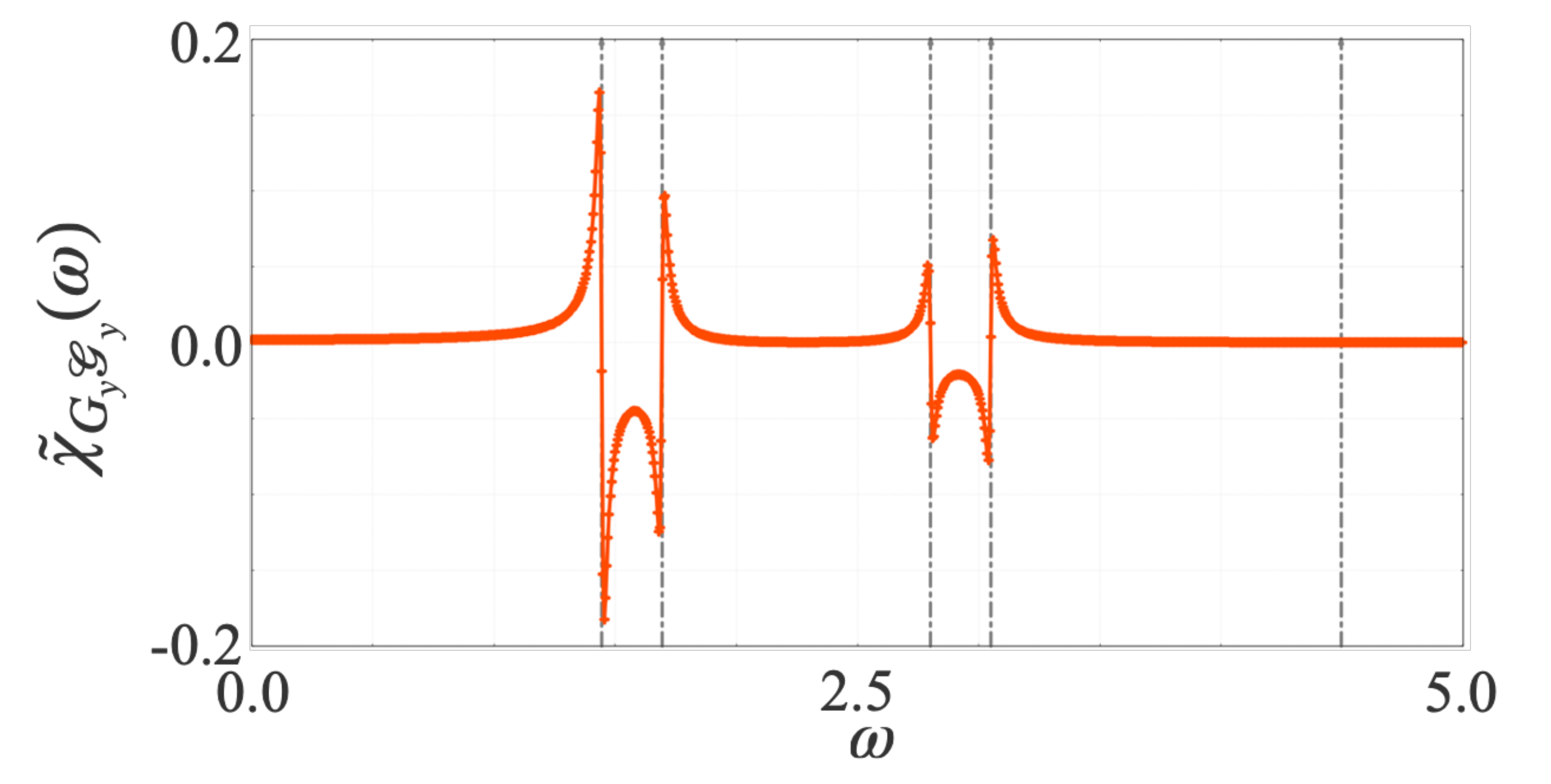}
  \caption{
  \label{f:chi_GydExdz}
  Frequency dependence of the real part of the susceptibility $\tilde{\chi}_{G_y \mathcal{G}_y}(\omega)$ for $r_{ba}=1$.
  Gray dashed lines indicate the resonant frequencies.
  }
\end{figure} 

First, we consider a minimal tight-binding model on a finite trapezoidal cluster with $C_{\rm 2v}$ symmetry, as illustrated in Fig.~\ref{f:model}(a). 
The trapezoidal geometry is characterized by the lengths $u_a$, $u_b$, and $u_h$ as shown in Fig.~\ref{f:model}(a), where we refer to the geometric asymmetry as the inequivalence between the top and bottom bases of the trapezoid.
The model consists of $s$-orbitals on six sites and incorporates the minimal ingredients required for ferroaxial responses, namely geometric asymmetry and spin--orbit coupling. 
The Hamiltonian is given as
\begin{align}
  \mathcal{H}_0 &= \mathcal{H}_t + \mathcal{H}_{\rm SOC},\\
  \mathcal{H}_t &= -\sum_{\langle i,j \rangle, \sigma} t_{ij} c_{i\sigma}^\dagger c_{j\sigma} + \text{H.c.}, \\
  \mathcal{H}_{\rm SOC} &= \sum_{\langle i,j \rangle, \sigma, \sigma'} c_{i \sigma}^\dagger \Lambda_{ij} {\sigma_y}_{\sigma\sigma'} c_{j \sigma'},
\end{align}
where $c_{i\sigma}^\dagger$ ($c_{i\sigma}$) is the creation (annihilation) operator of an electron with spin $\sigma$ at site $i$, and $t_{ij}$ is the hopping amplitude between the nearest-neighbor sites $i$ and $j$.
The Hamiltonian $\mathcal{H}_t$ represents the hopping term, where we set $t_{12}=t_{34}=t_a, t_{35}=t_{46}=t_b$, and $t_{13}=t_{24}=t_h$.
The Hamiltonian $\mathcal{H}_{\rm SOC}$ represents the spin--orbit coupling term, where $\Lambda_{ij}$ denotes its matrix elements.
Based on the symmetry considerations, we retain the $y$-spin component, with $\Lambda_{13}= -\Lambda_{24} = -\Lambda_{31} = \Lambda_{42} = -i/2\sqrt{2} \lambda$, where $\lambda$ is the strength of the spin--orbit coupling. 
The energy unit is set to be $t_a = t_b = t_h = 1$. 
We neglect the dependence of the hopping amplitudes on the lattice parameters $u_a$, $u_b$, and $u_h$ shown in Fig.~\ref{f:model}(a), for simplicity. 
The energy levels of the Hamiltonian $\mathcal{H}_0$, which is measured from the lowest energy level, are shown in Fig.~\ref{f:chi_GydExdz} as gray dashed lines.

We next introduce the electric field gradient induced by the geometric structure.
Hereafter, we set $u_a=u_h$ for simplicity. 
As shown in Fig.~\ref{f:model}(b), applying a voltage across the trapezoidal legs generates spatially varying electric fields.
Since the electrostatic potential at each site depends on its distance from the electrode, these spatial variations are incorporated as a site-dependent potential as 
\begin{align}
  \mathcal{H}_{\rm grad} 
  &= {\rm Diag}[1, -1, 1/(2r_{ba}+1), -1/(2r_{ba}+1), 1, -1]\notag\\
  &\times qE_0 u_a /2\label{eq:H_grad}
  \end{align}
where $E_0$ is the electric field strength, $q$ is the elementary charge, and $r_{ba}\equiv u_b/u_a$; we set $q=1$. 
The diagonal form of Eq.~\eqref{eq:H_grad} represents the electrostatic potential originating from the distance to the electrodes and consequently produces a spatially nonuniform electric field, namely $E_x$ varying along $z$ and $E_z$ varying along $x$. 
For example, $\mathcal{H}_{\rm grad}/(qE_0)$ can be decomposed as $p_x^{\rm (top)}+[1/\{2(1+2r_{ba})\}] p_x^{\rm (bottom)}$, where $p_x^{\rm (top)}$ and $p_x^{\rm (bottom)}$ denote the $x$ component of electric polarization on the top and bottom bases, respectively.
Meanwhile, the limit $r_{ba} \to 0$ corresponds to the disappearance of the geometric asymmetry.

These spatial variations of the electric field can be decomposed into symmetric and antisymmetric components in the continuum description as
\begin{align}
  \tilde{E}^{\rm (S)}_{zx} &\equiv
  \frac{\partial E_z(\bm{r})}{\partial x}
  +
  \frac{\partial E_x(\bm{r})}{\partial z},\\
  \tilde{E}^{\rm (A)}_y &\equiv
  \frac{\partial E_z(\bm{r})}{\partial x}
  -
  \frac{\partial E_x(\bm{r})}{\partial z}.
\end{align}
Accordingly, Eq.~\eqref{eq:H_grad} can be rewritten as
\begin{align}
  \mathcal{H}_{\rm grad}
  = c_1 \tilde{E}^{\rm (S)}_{zx}
  + c_2 \tilde{E}^{\rm (A)}_y
  \equiv \mathcal{G}_y,
\end{align}
where $c_1$ and $c_2$ represent the weights of the symmetric and antisymmetric components, respectively.
The antisymmetric component $\tilde{E}^{\rm (A)}_y$ corresponds to the $y$ component of the curl of the electric field, i.e., $(\nabla \times \bm{E})_y$, and hence acts as the conjugate field to the ferroaxial moment $G_y$.

From the symmetry viewpoint, the system possesses $D_{\rm 2h}$ symmetry in the absence of geometric asymmetry, which is reduced to $C_{\rm 2v}$ by the asymmetric geometry. 
Within the $C_{\rm 2v}$ point group, the out-of-plane ferroaxial moment $G_y$ belongs to the $B_1$ irreducible representation of the $C_{\rm 2v}$ point group. 
In the presence of an electric field gradient, the symmetry of the system lowers from $C_{\rm 2v}$ to $C_{\rm s}$, under which the ferroaxial moment transforms as the identity irreducible representation.
This symmetry reduction activates ferroaxial responses that are otherwise constrained in the uniform-field limit.

The lowest-order contribution to the out-of-plane ferroaxial moment induced by the electric field gradient is expressed as 
\begin{align}
  G_y(\omega) = \chi_{G_y \mathcal{G}_y}(\omega)\mathcal{G}(\omega).
\end{align}
Here, the operator expression of $G_y$ is represented microscopically by spin-dependent imaginary hoppings [Fig.~\ref{f:model}(b)], using the open-source Python library, MultiPie as follows~\cite{Kusunose_PhysRevB.107.195118}: 
  \begin{align}
    \hat{G}_y = \sum_{\langle i,j \rangle, \sigma, \sigma'} c_{i \sigma}^\dagger g_{ij} {\sigma_y}_{\sigma\sigma'} c_{j \sigma'},
  \end{align}
where $g_{ij}$ is the matrix element given by $g_{13}= g_{24} = -g_{31} = -g_{42} = -i/2\sqrt{2}$ and zero for other $i$ and $j$.
This form provides an alternative description of the ferroaxial moment in terms of cross-product-type spin--orbit coupling~\cite{hayami2018prb_Classification_of_atomic-scale_multipoles}, with the same symmetry as $\sum_i\bm{r}_i\times \bm{p}_i$. 
The coefficient $\chi_{G_y \mathcal{G}_y}(\omega)$ is the susceptibility of $G_y$, which is evaluated as within the linear response theory
\begin{align}
  \chi_{G_y \mathcal{G}_y}(\omega) 
  =
  -\sum_{ij} [G_y]_{ij} [\mathcal{G}_y]_{ji} 
\frac{f_i-f_j}{-\hbar\omega + i\hbar\delta +(\xi_i-\xi_j)},\label{eq:chi_GydExdz}
\end{align}
where $\xi_i$ is the eigenvalue of the Hamiltonian $\mathcal{H}_0$ and $f_i\equiv f(\xi_i)$ is the Fermi distribution function.
$[G_y]_{ij}=\langle i | G_y | j \rangle$ and $[\mathcal{G}_y]_{ji}=\langle j | \mathcal{G}_y | i \rangle$ are the matrix elements of the ferroaxial-moment and the electric-field-gradient operators between the eigenstates $|i\rangle$ and $|j\rangle$, respectively.
We set $\hbar=1$ hereafter. 
The parameter $\delta$ denotes the scattering rate, and we set $\delta=0.01$ and the temperature $T=0.1$ in the following calculations.

We note that, under $C_{\rm 2v}$ symmetry, the ferroaxial moment is induced not only by the electric field gradient but also by a uniform electric field $E_x$, since $E_x$ belongs to the same irreducible representation as $G_y$.
To isolate the contribution originating purely from the electric field gradient, we define the subtracted susceptibility as
\begin{align}
  \tilde{\chi}_{G_y \mathcal{G}_y}(\omega, r_{ba})
  \equiv 
  \chi_{G_y \mathcal{G}_y}(\omega, r_{ba})
  -
  \chi_{G_y \mathcal{G}_y}(\omega, r_{ba}=0),
\end{align}
which removes the uniform-field contribution. 
This contribution vanishes in the $D_{\rm 2h}$ limit, where the lower base angle is a right angle.

Figure~\ref{f:chi_GydExdz} shows the frequency dependence of the real part of $\tilde{\chi}_{G_y \mathcal{G}_y}(\omega)$ for $r_{ba}=1$ at filling $n = 2/12$, corresponding to occupation of the lowest energy level including spin degeneracy. 
The behavior of $\tilde{\chi}_{G_y \mathcal{G}_y}(\omega)$ shows that the ferroaxial moment is induced by the electric field gradient and is enhanced around the resonant frequencies.
In particular, the response is maximized near the resonance associated with the HOMO--LUMO gap, where the energy denominator in Eq.~(\ref{eq:chi_GydExdz}) is minimized. 
These results confirm that the electric field gradient acts as the conjugate field of ferroaxial moments, consistent with the symmetry argument above.

To further identify the important model parameters responsible for activating $G_y$, we perform the expansion analysis following Refs.~\onlinecite{hayami2020prb_bottom-up, R.Oiwa2022jpsj_nonl-respo}. 
As a result, we find that the spin--orbit coupling $\lambda$ necessarily appears in the expansion of $\tilde{\chi}_{G_y \mathcal{G}_y}(\omega, r_{ba})$, indicating that spin--orbit coupling is essential for inducing the spin-dependent ferroaxial moment under the electric field gradient. 
Moreover, we also find that each expansion term is proportional to $r_{ba}/(2r_{ba}+1)$, highlighting the crucial role of geometric asymmetry.
This scaling behavior will be further discussed in Sec.~\ref{sec:slope_dependence}.

\section{Transverse magnetization in the presence of electric field gradient}
\label{sec:TM}

We next investigate the TM induced by a magnetic field in the presence of an electric field gradient. 
In ferroaxial systems, unconventional transverse responses proportional to the cube of the magnetic field have been shown~\cite{inda2023jpsj}. 
Here, we demonstrate that the electric field gradient activates a distinct mechanism, leading to a TM that is linear in both the magnetic field and the electric field.

To this end, we introduce the time-dependent magnetic field oscillating at frequency $\omega$ as the Zeeman coupling
\begin{align}
  \mathcal{H}_{\rm Zeeman} = - \sum_i \bm{H}(t) \cdot \bm{s}_i ,
\end{align}
where $\bm{s}_i$ is the spin operator at site $i$.
In the absence of the electric field, the spin magnetization induced by the external magnetic field can be Fourier transformed and expanded as follows: 
\begin{align}
  & 
  M^{{\rm s}(E=0)}_\eta(\omega) \notag\\
  &=
  \sum_{\mu}
  \chi_{s_\eta H_\mu}(\omega) H_\mu(\omega)
  +
  \Bigg[\sum_{\mu\nu\kappa}
  \int\int\int \frac{d\omega_1 d\omega_2 d\omega_3}{(2\pi)^2}\notag\\
  &\times \chi_{s_\eta H_\mu H_\nu H_\kappa}(\omega; \omega_1, \omega_2, \omega_3) H_\mu(\omega_1)H_\nu(\omega_2)H_\kappa(\omega_3)\Bigg] + \cdots \label{eq:M_s_E0},
\end{align}
where $M^{\rm s}_\eta\equiv (1/N)\sum_{i}^N \langle s_\eta \rangle$ represent the spin magnetization, $\chi_{s_\eta H_\mu}(\omega)$ and $\chi_{s_\eta H_\mu H_\nu H_\kappa}(\omega; \omega_1, \omega_2, \omega_3)\equiv \delta(\omega-\omega_1-\omega_2-\omega_3) \chi_{s_\eta H_\mu H_\nu H_\kappa}(\omega_1, \omega_2, \omega_3)$ for $\eta, \mu,\nu, \kappa = x,y,z$ represent the linear and third-order magnetic susceptibilities, respectively; even-order contributions are forbidden in time-reversal-even systems. 
Under $C_{\rm 2v}$ symmetry, only $\chi_{s_\eta H_\eta}(\omega)$, $\chi_{s_\eta H_\eta H_\eta H_\eta}(\omega_1, \omega_2, \omega_3)$, and $\chi_{s_\eta H_\mu H_\mu H_\eta}(\omega_1, \omega_2, \omega_3)$ for $\eta, \mu = x,y,z$ can be finite.
In particular, under the magnetic field along the $z$ direction, the response is limited to the longitudinal magnetization parallel to the applied magnetic field; no transverse magnetizations perpendicular to the magnetic field are induced.

In the presence of an electric field or its spatial gradient, additional response channels, including the TM, become active. 
Taking into account the time-reversal symmetry breaking due to the dissipation under the electric field, the magnetization acquires additional contributions that can be expressed as follows:
\begin{align}
  & M^{{\rm s}^{(E \neq 0)}}_\eta(\omega)\notag  \\
  &=  
\Bigg[
  \sum_{\mu\nu}
  \int\int \frac{d\omega_1 d\omega_2}{2\pi} 
  \chi_{s_\eta H_\mu E_\nu}(\omega; \omega_1, \omega_2) H_\mu(\omega_1)E_\nu(\omega_2)\Bigg]\notag\\
  &+\Bigg[
  \sum_{\mu\nu}
  \int\int \frac{d\omega_1 d\omega_2}{2\pi} 
  \chi_{s_\eta H_\mu \tilde{E}^{\rm (A)}_\nu}(\omega; \omega_1, \omega_2) H_\mu(\omega_1)\tilde{E}^{\rm (A)}_\nu(\omega_2)\Bigg]\notag\\
  &+\Bigg[
  \sum_{\mu\nu\kappa}
  \int\int \frac{d\omega_1 d\omega_2}{2\pi} 
  \chi_{s_\eta H_\mu \tilde{E}^{\rm (S)}_{\nu\kappa}}(\omega; \omega_1, \omega_2) H_\mu(\omega_1)\tilde{E}^{\rm (S)}_{\nu\kappa}(\omega_2)\Bigg]\notag\\
  &+\Bigg[
  \sum_{\mu\nu}
  \int\int \frac{d\omega_1 d\omega_2}{2\pi} 
  \chi_{s_\eta H_\mu H_\nu}(\omega; \omega_1, \omega_2) H_\mu(\omega_1)H_\nu(\omega_2)\Bigg]+ \cdots\label{eq:M_s_E_nonzero},
\end{align}
where $\chi_{s_\eta H_\mu E_\nu}(\omega; \omega_1, \omega_2)$ represents the contribution from the uniform electric field, while $\chi_{s_\eta H_\mu \tilde{E}^{\rm (A)}_\nu}(\omega; \omega_1, \omega_2)$ and $\chi_{s_\eta H_\mu \tilde{E}^{\rm (S)}_{\nu\kappa}}(\omega; \omega_1, \omega_2)$ arise from the electric field gradient.
Another contribution, $\chi_{s_\eta H_\mu H_\nu}(\omega; \omega_1, \omega_2)$, also arises when the electric field is applied. This contribution is induced by time-reversal symmetry breaking originating from the dissipation, but is neglected as it is irrelevant to the following analysis in terms of the TM. 
Hereafter, we omit the frequency dependence of the susceptibilities when it is clear from the context.

Among the additional electric-field-induced contributions, transverse components are generated by the electric field gradient. 
Specifically, under $C_{2\rm v}$ symmetry, nonzero transverse tensor components are given by $\chi_{s_x H_y \tilde{E}^{\rm (A)}_z}$, $\chi_{s_x H_z \tilde{E}^{\rm (A)}_y}$, $\chi_{s_y H_x \tilde{E}^{\rm (A)}_z}$, $\chi_{s_y H_z \tilde{E}^{\rm (A)}_x}$, $\chi_{s_z H_x \tilde{E}^{\rm (A)}_y}$, $\chi_{s_z H_y \tilde{E}^{\rm (A)}_x}$, and $\chi_{s_\eta H_\mu \tilde{E}^{\rm (S)}_{\mu\eta}}$ for $\eta, \mu =  x,y,z$. 
This indicates that the TM is induced by the combined effect of the magnetic field and the electric field gradient.
It is noted that $\chi_{s_\eta H_\mu \tilde{E}^{\rm (S)}_{\nu\kappa}}$ also gives the longitudinal contribution represented by $\chi_{s_\eta H_\eta \tilde{E}^{\rm (S)}_{\eta\eta}}$.

In contrast, all components of $\chi_{s_\eta H_\mu E_\nu}(\omega; \omega_1, \omega_2)$ vanish under $D_{\rm 2h}$ symmetry, i.e., in the absence of geometric asymmetry ($r_{ba}\rightarrow 0$).
Thus, unlike the electric-field-gradient-induced contributions discussed above, these uniform-field responses are not associated with the geometry-induced field-gradient effect considered in this work. 
Accordingly, we neglect them in the following and focus on the TM induced by the electric field gradient. 
The neglected uniform-field contribution is shown in Appendix~\ref{appendix:w_uniform_E}.

A key consequence is that the induced TM is linear in both the magnetic field and the electric field gradient, in contrast to the cubic magnetic-field dependence expected in ferroaxial ordered states~\cite{inda2023jpsj}. 
This linearity implies that the response can be tuned by varying either the magnetic field or the electric field gradient.
Notably, while both longitudinal and transverse magnetizations scale linearly with the magnetic field, the presence of the electric field gradient enables a finite transverse response without being suppressed relative to the longitudinal one. 
This provides a clear advantage for experimental detection. 
Moreover, the TM exhibits an antisymmetric dependence on the magnetic field within the present model: $\chi_{s_x H_z \tilde{E}^{\rm (A)}_y}(\omega; \omega_1, \omega_2) = -\chi_{s_z H_x\tilde{E}^{\rm (A)}_y}(\omega; \omega_1, \omega_2)$ and $\chi_{s_x H_z \tilde{E}^{\rm (S)}_{zx}}=-\chi_{s_z H_x \tilde{E}^{\rm (S)}_{zx}}$, since both the input and output operators are described by the Pauli matrices satisfying the anticommutation relations. 
Such an antisymmetric linear response is prohibited by Onsager reciprocity in the absence of the electric field gradient, and thus represents a unique signature of responses induced by the electric field gradient.

To quantitatively evaluate the TM under the electric field gradient, we perform real-time numerical simulations for the trapezoidal model with time-dependent electromagnetic fields.
We adopt the Peierls substitution to incorporate the effect of the external electric field $\bm{E}$ as 
\begin{align}
t'_{ij} \rightarrow t'_{ij} \exp\left[i\int_{\bm{r}_i}^{\bm{r}_j} \bm{A}(t, \bm{r}) \cdot d\bm{r}\right],
\end{align} 
where $t'_{ij}= t_{ij}, \Lambda_{ij}, g_{ij}$ and $\bm{A}(t, \bm{r})$ is the spatially inhomogeneous vector potential satisfying $\bm{E}(t)=-\partial \bm{A}(t, \bm{r})/\partial t$. 
Its spatial dependence encodes the geometry-induced electric field gradient.
We decompose the vector potential as $A(t,\bm{r}) \equiv |\bm{A}(t, \bm{r})| = A_t(t) A_r(\bm{r})$, separating temporal and spatial variations. 
The spatial profile $A_r(\bm{r})$ is chosen as $A_r(\bm{r})=1$ for $t_{12}$, $A_r(\bm{r})=1/(2r_{ba}+1)$ for $t_{34}$, and $A_r(\bm{r})=r_{ba}/(2r_{ba}+1)$ for $t_{13}$, $t_{24}$, $t_{35}$, and $t_{46}$.

\begin{figure}[t!]
  \centering
  \includegraphics[width=\linewidth]{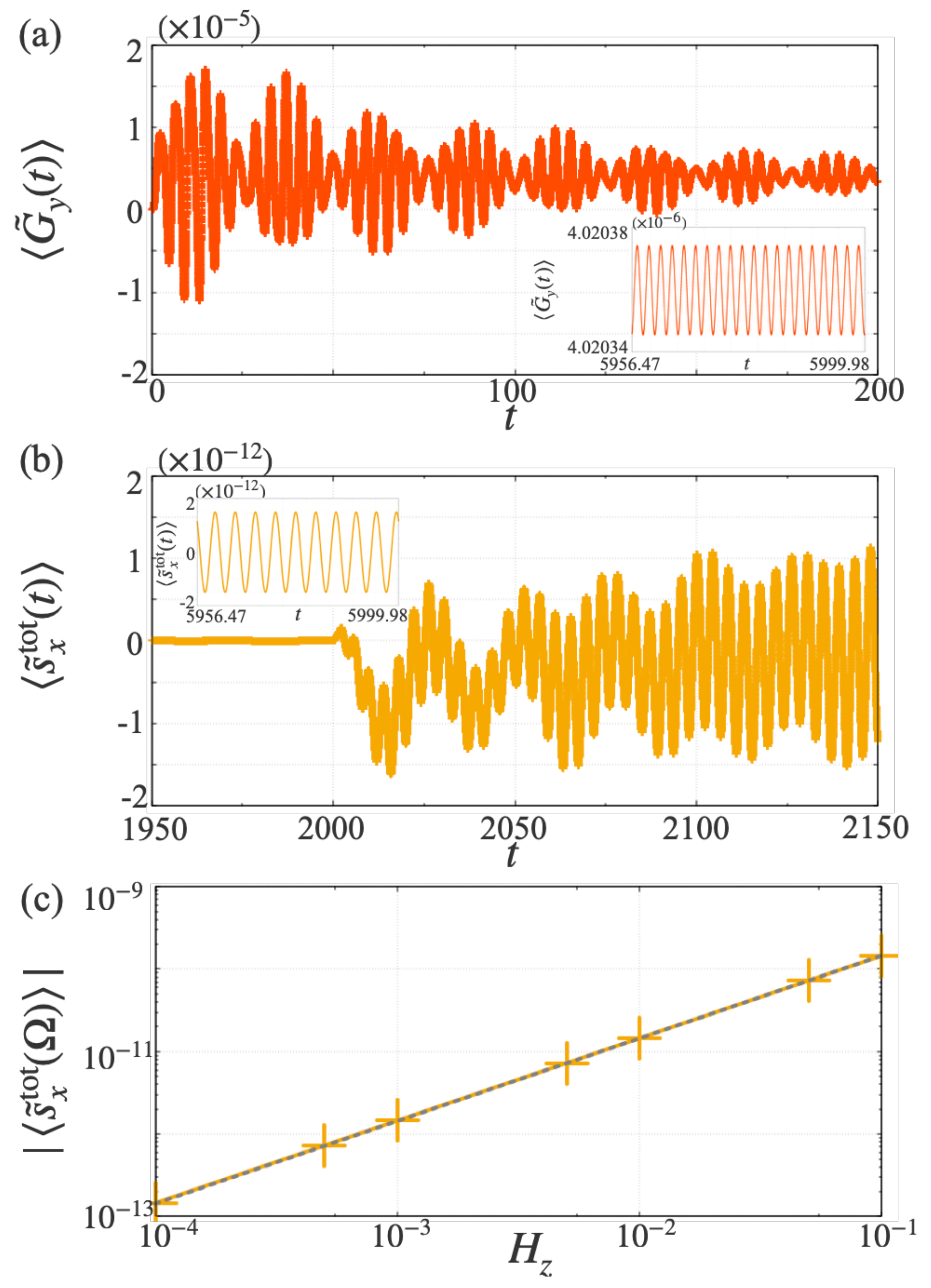}
  \caption{
  \label{f:Gy_sx_tdep}
  Time evolution of (a) the ferroaxial moment $\langle \tilde{G}_y (t)\rangle$ and (b) the transverse magnetization $\langle \tilde{s}_x^{\rm tot}(t) \rangle$ under the DC electric field and oscillating magnetic field. 
  The insets of (a) and (b) show the time evolution of $\langle \tilde{G}_y (t)\rangle$ and $\langle \tilde{s}_x^{\rm tot}(t) \rangle$  in the interval $5956.47 \leq t \leq 5999.98$, respectively.
  The parameters are set to $H_z=1.0 \times 10^{-3}$, $E_0=1.0 \times 10^{-3}$, $\lambda = 0.1$, $\delta = 0.01$, and $\Omega=1.44372$, which corresponds to the resonant frequency of the HOMO--LUMO gap.
  (c) Magnetic field dependence of the transverse magnetization at $\omega=\Omega$ under the DC electric field.
  The gray dashed line shows a fit to $ax^b$ with $a=1.46862\times 10^{-9}$ and $b= 1.00323$.
  }
\end{figure} 

\begin{figure}[t!]
  \centering
  \includegraphics[width=\linewidth]{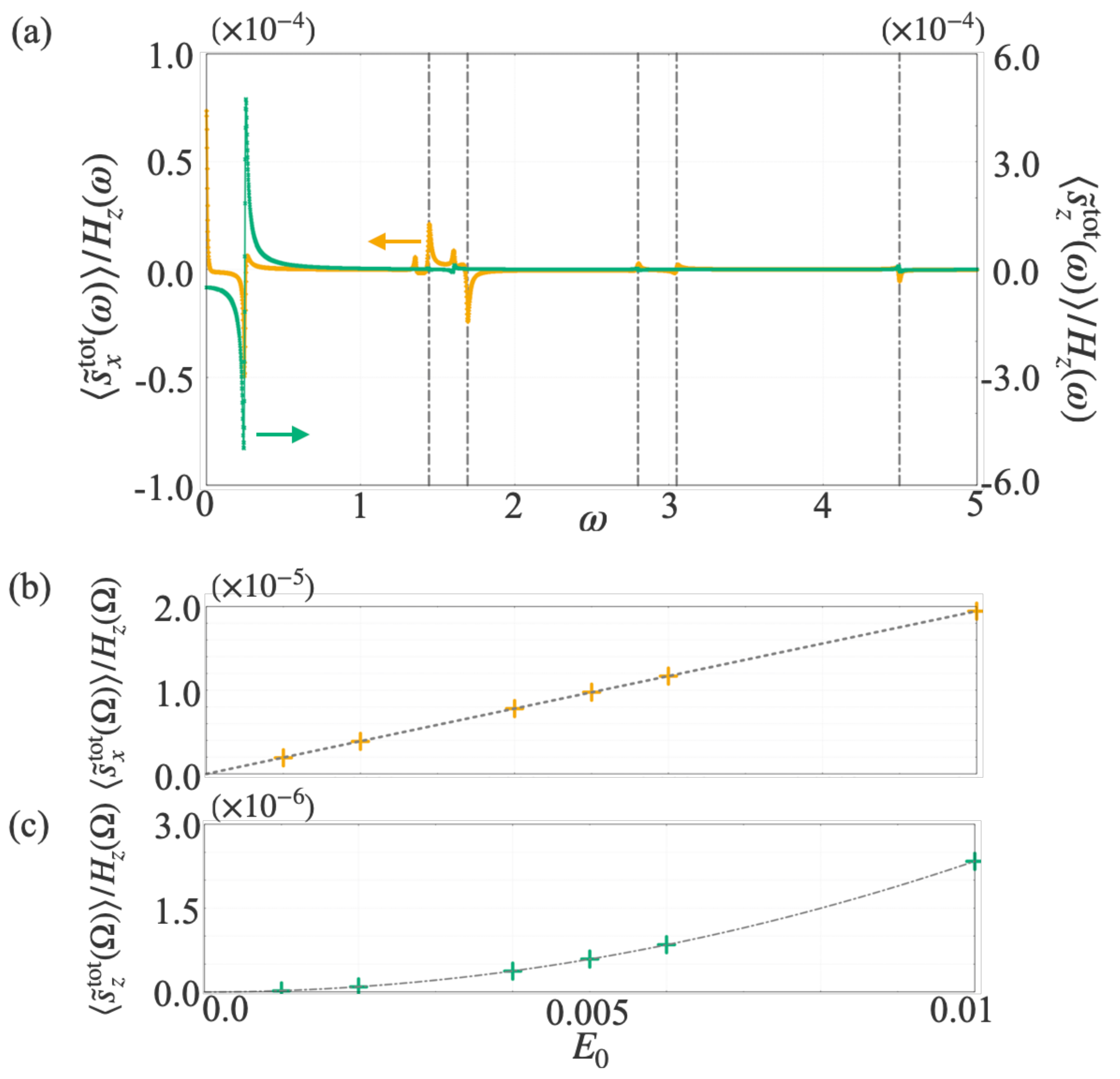}
  \caption{
  \label{f:tra_longi_mag}
  (a) Frequency dependence of the transverse magnetic response $\langle \tilde{s}_x^{\rm tot}(\omega)\rangle/H_z(\omega)$ and longitudinal magnetic response $\langle \tilde{s}_z^{\rm tot}(\omega)\rangle/H_z(\omega)$ induced by a pulsed magnetic field along $z$ axis under the AC electric field. 
  We set the frequency of the electric field as $\Omega=1.44372$.
  (b) Electric field dependence of the transverse magnetic response at $\Omega=1.44387$.
  The gray dashed line shows a fit to $ax^b$ with $a=0.0189767$ and $b=0.996497$.
  (c) Electric field dependence of the longitudinal magnetic response at $\Omega=1.44387$.
  The gray dashed line shows a fit to $ax^b$ with $a=2.25483$ and $b=1.99463$.
  }
\end{figure}

The time evolution of the system is described by the single-particle density matrix $\rho_{ij}(t) = \langle c_i^\dagger c_j \rangle(t) \equiv {\rm Tr}[c_i^\dagger c_j\rho(t)]$, where $\rho(t)$ is the density matrix at time $t$.
It is calculated by solving the von Neumann equation~\cite{Yue:22}:
\begin{align}
  \frac{d\rho(t)}{dt} = -i [\mathcal{H}(t), \rho(t)] - \gamma \{\rho(t) - \rho_{\rm eq}(t)\}, \label{eq:vonNeumann}
\end{align}
where $\rho_{\rm eq}(t)$ is the single-particle density matrix in equilibrium: $\rho_{\rm eq}(t) = \sum_i f_i |i\rangle\langle i|$.
We incorporate relaxation effects phenomenologically using the relaxation-time approximation in the second term of Eq.~(\ref{eq:vonNeumann})~\cite{PhysRevB.110.125111,hattori2024effect,xsrf-t1hj,qsxr-c2pq,PhysRevB.106.035204}; the relaxation rate is given by $\gamma = 1/\tau$, which corresponds to $\delta$ in Eq.~(\ref{eq:chi_GydExdz}).
We compute the time evolution of $\rho(t)$ using the fourth-order Runge-Kutta method and evaluate the total magnetization: $s_\mu^{\rm tot} = \sum_{i} s_{\mu,i}$ corresponding to $N ({M^{\rm s}}^{(E=0)}_\mu+ {M^{\rm s}}^{(E\neq 0)}_\mu)$. 

Figures~\ref{f:Gy_sx_tdep}(a) and (b) show the time evolution of $G_y$ and the TM, respectively, under a DC electric field and an oscillating magnetic field. 
Focusing on the contributions from the electric field gradient, we evaluate the ferroaxial moment and the TM purely originating from the electric field gradient as
\begin{align}
  \langle \tilde{G}_y(t, r_{ba}) \rangle
  &\equiv 
  \langle G_y(t, r_{ba}) \rangle - \langle G_y(t, r_{ba}=0) \rangle,\\
  \langle \tilde{s}^{\rm tot}_x(t, r_{ba}) \rangle
  &\equiv 
  \langle s_x^{\rm tot}(t, r_{ba}) \rangle - \langle s_x^{\rm tot}(t, r_{ba}=0) \rangle.
\end{align}
The electric field is applied as $\bm{A}_t(t) = (-E_0 t,0 ,-E_0 t)$ for $0 \leq t \leq 6000$ with $E_0=1.0 \times 10^{-3}$, while the magnetic field $H_z(t) = h_z \cos(\Omega t)$ is applied along the $z$ direction in the inverval $2000\leq t \leq 6000$ with $H_z=1.0 \times 10^{-3}$. 
We choose $\Omega=1.44372$, corresponding to the HOMO--LUMO resonant frequency.
As discussed in Sec.~\ref{sec:kubo}, the ferroaxial moment $G_y$ is generated simultaneously with the application of the electric field [Fig.~\ref{f:Gy_sx_tdep}(a)]. 
In contrast, Fig.~\ref{f:Gy_sx_tdep}(b) demonstrates that the TM is induced only when the magnetic field is applied in the presence of the electric field.

After the system reaches a nonequilibrium steady state, we evaluate the magnetization from the expectation value of the spin operator. 
To obtain the frequency-dependent spin moment $\langle \tilde{s}_x^{\rm tot}(\omega)\rangle$, we perform the Fourier transformation of $\langle \tilde{s}_x^{\rm tot}(t)\rangle$ over the time window $5956.47 \leq t \leq 5999.98$, as shown in the inset of Fig.~\ref{f:Gy_sx_tdep}(b), following previous studies~\cite{PhysRevLett.127.127402, PhysRevB.109.064407, hattori2025nonlinear, hattori2025dirac, inda2026antiparallel}. 
The resulting $h_z$ dependence of $\langle \tilde{s}_x^{\rm tot}(\Omega)\rangle$ is shown in Fig.~\ref{f:Gy_sx_tdep}(c), where the TM is linearly proportional to the magnetic field.
We also confirm that the TM arises from an antisymmetric component, consistent with the anticommutation relations of Pauli matrices as discussed above.

To further analyze the frequency dependence of the TM, we apply an AC electric field described by $\bm{A}_t(t)= (-E_0 \sin(\Omega t)/\Omega, 0, -E_0 \sin(\Omega t)/\Omega)$ together with a magnetic-field pulse $H_z(t)= h_z \exp[-(t-t_0)^2/(2\sigma^2)]$, with $E_0= h_z = 1.0 \times 10^{-3}$, $\Omega=1.44372$, $t_0=0.2$, and $\sigma=0.03$. 
Figure~\ref{f:tra_longi_mag}(a) shows the frequency dependence of the TM response $\langle \tilde{s}^{\rm tot}_x(\omega)\rangle /H_z(\omega)$ and the longitudinal magnetic response $\langle \tilde{s}^{\rm tot}_z(\omega)\rangle /H_z(\omega)$ under the electric field gradient; $H_z(\omega)$ is the Fourier transform of $H_z(t)$. 
A comparison of the peak values reveals that the maximum of the induced TM reaches approximately 18 percent of the longitudinal one. 
We also find a qualitative difference between the two responses:
the TM exhibits peaks at the resonant frequencies, whereas the longitudinal magnetization changes sign across the resonance.
This result indicates that the TM can exceed the longitudinal magnetization near the resonant frequencies.

Moreover, we confirm that the TM is linear in the electric field, whereas the longitudinal magnetization shows a quadratic dependence on the electric field, as shown in Figs.~\ref{f:tra_longi_mag}(b) and (c), where we set $\Omega=1.44372$. 
Although a local longitudinal magnetization can be induced as a linear response to the uniform electric field $E_z$ under $C_{\rm 2v}$ symmetry, the contributions from the left- and right-hand sides of the system cancel each other, yielding a vanishing net linear response. 
These results are consistent with the symmetry analysis and demonstrate that the TM can be effectively enhanced relative to the longitudinal magnetization by tuning the frequency and amplitude of the electric field.

\section{Scaling of ferroaxial and transverse magnetic responses with leg inclination}
\label{sec:slope_dependence}
\begin{figure}[t!]
  \centering
  \includegraphics[width=\linewidth]{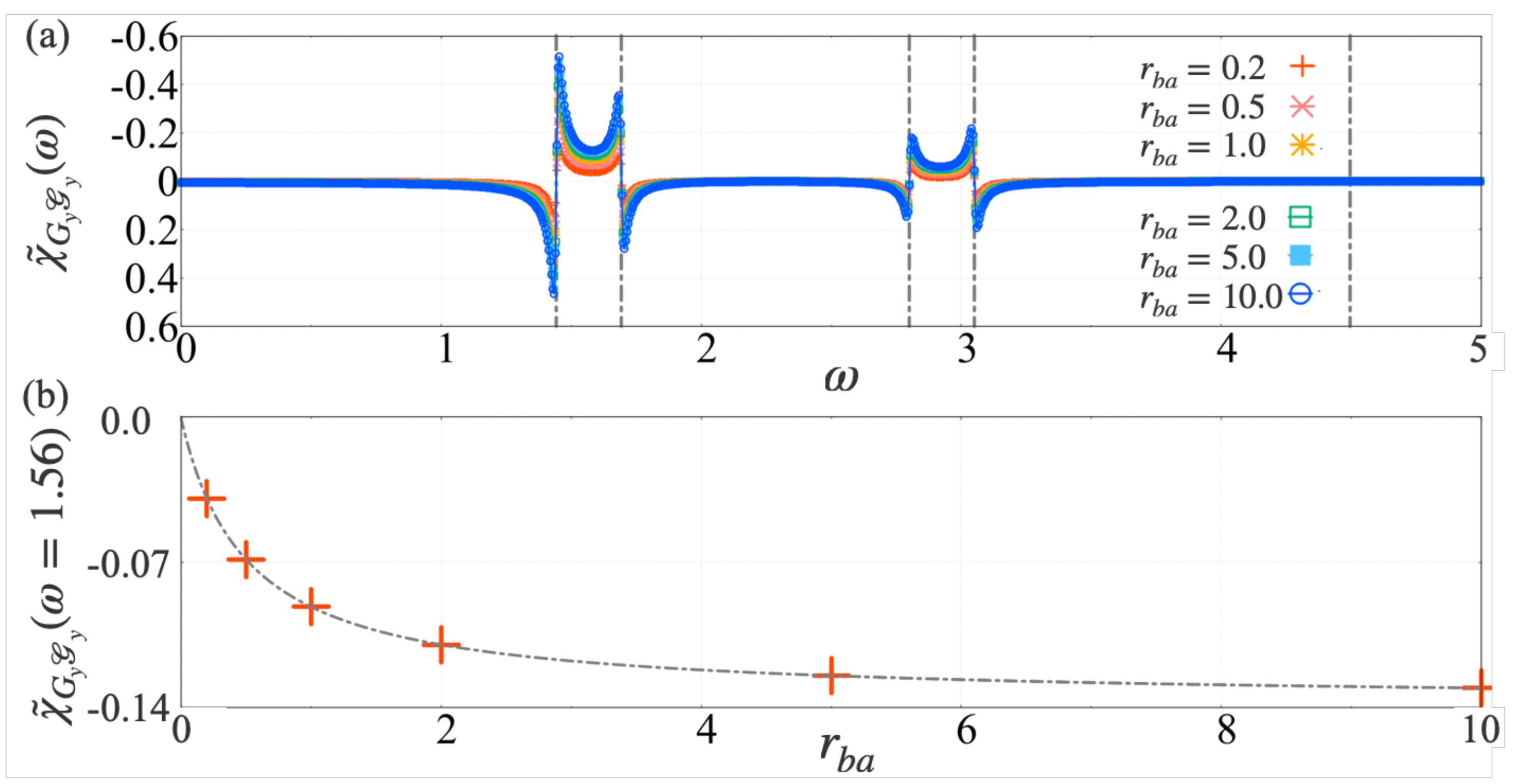}
  \caption{
  \label{f:ratio_dep_Gy}
  (a) Frequency dependence of the susceptibility $\tilde{\chi}_{G_y \mathcal{G}_y}(\omega)$ for several values of $r_{ba}$.
  (b) Leg-inclination dependence of $\tilde{\chi}_{G_y \mathcal{G}_y}(\omega)$ at $\omega=1.56$, showing a scaling behavior consistent with $\alpha r_{ba}/(\beta r_{ba}+1)$.
  The gray dashed line denotes a fit with $\alpha=-0.274051$ and $\beta=2$.
  Gray dashed lines indicate the resonant frequencies.
  }
\end{figure}
\begin{figure}[t!]
  \centering
  \includegraphics[width=\linewidth]{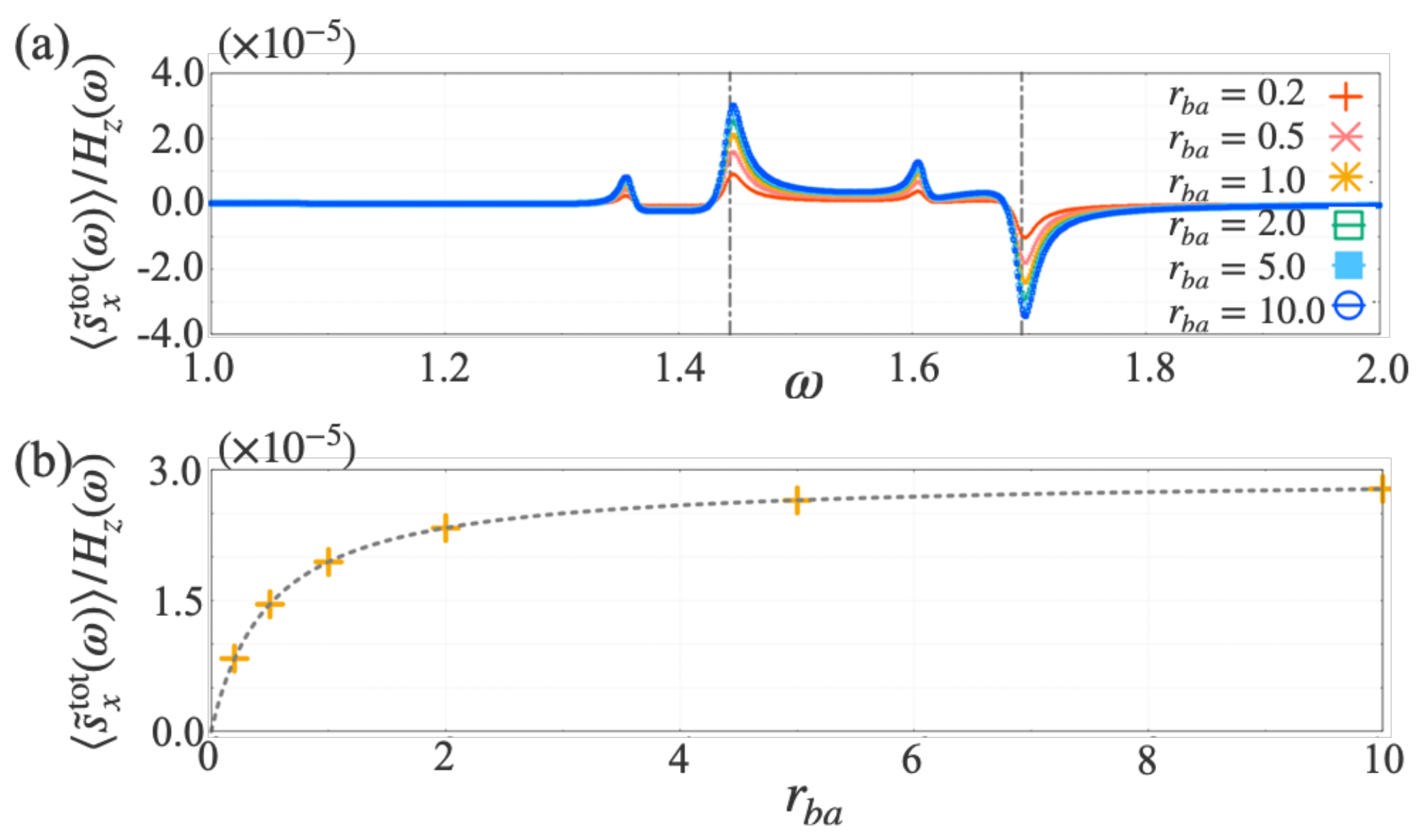}
  \caption{
  \label{f:ratio_dep_sx}
  (a) Frequency dependence of the transverse magnetic response $\langle \tilde{s}_x^{\rm tot}(\omega)\rangle /H_z(\omega)$ for several values of $r_{ba}$.
  Gray dashed lines indicate the resonant frequencies.
  (b) Leg-inclination dependence of $\langle \tilde{s}_x^{\rm tot}(\omega)\rangle /H_z(\omega)$ at $\omega=1.44387$, showing a universal scaling consistent with $\alpha r_{ba}/(\beta r_{ba}+1)$.
  The gray dashed line represents a fit with $\alpha=5.82202\times 10^{-5}$ and $\beta=1.99544$.
  }
\end{figure}

Finally, we investigate the influence of the leg inclination of the trapezoidal geometry, which controls the strength of the electric field gradient, on both the ferroaxial moment and the TM. 
We adopt the same parameters as those used in Fig.~\ref{f:chi_GydExdz} for the susceptibility and those used in Fig.~\ref{f:tra_longi_mag} for the TM.

Figure~\ref{f:ratio_dep_Gy}(a) shows $r_{ba}$ dependence of the ferroaxial susceptibility $\tilde{\chi}_{G_y \mathcal{G}_y}$. 
The susceptibility increases monotonically with $r_{ba}$, indicating that the ferroaxial response is systematically enhanced as the leg inclination becomes smaller.
This behavior is consistent with the expansion analysis in Sec.~\ref{sec:kubo}, where the contribution scales with the geometric factor $r_{ba}/(2r_{ba}+1)$ [see also Fig.~\ref{f:ratio_dep_Gy}(b)]. 
The monotonic increase reflects the enhancement of the electric field gradient arising from geometric asymmetry.

A qualitatively similar behavior is observed for the TM induced by the electric field gradient, as shown in Fig.~\ref{f:ratio_dep_sx}(a).
For both the ferroaxial moment and the TM, the enhancement persists over a wide frequency range. 
Moreover, as shown in Figs.~\ref{f:ratio_dep_Gy}(b) and \ref{f:ratio_dep_sx}(b), the $r_{ba}$ dependence of both quantities collapses onto a common scaling behavior proportional to $r_{ba}/(2r_{ba}+1)$.
This universal scaling suggests that both the ferroaxial response and the TM share a common microscopic origin, namely the electric field gradient controlled by the system geometry.

To gain further insight into the origin of this scaling, we evaluate the electric-field gradient at the center of the system.
Because the electric field is introduced only on the bonds through the spatially dependent vector potential, as described in Sec.~\ref{sec:TM}, we set the origin at the center of the system and approximately evaluate the conjugate field of the ferroaxial moment, corresponding to the electric-field gradient $(\nabla \times \bm{E})_y|_{\bm{r}=\bm{0}}$, as follows:
\begin{align}
  \frac{\partial E_z }{\partial x} \Big|_{\boldsymbol{r}=\boldsymbol{0}}
  &\sim 
  \frac{\frac{V_3 -V_1}{u_h} - \frac{V_4 - V_2}{u_h}}{u_a}\notag\\
  &=
  -\frac{2u_b V}{u_au_h(u_a+2u_b)},\notag\\
  \frac{\partial E_x}{\partial z} \Big|_{\boldsymbol{r}=\boldsymbol{0}}
  &\sim 
  \frac{\frac{V_2 -V_1}{u_a} - \frac{V_6 - V_5}{u_b}}{u_h}\notag\\
  &=
  \frac{2u_b V}{u_au_h(u_a+2u_b)},\notag\\
  \therefore (\nabla \times \bm{E})_y|_{\boldsymbol{r}=\boldsymbol{0}} 
  &\sim -\frac{4 V r_{ba}}{u_au_h(2r_{ba}+1)},
\end{align}
where $V_i (i=1,2,3,4,5,6)$ is the electrostatic potential at site $i$. 
This expression reproduces the same $r_{ba}/(2r_{ba}+1)$ dependence observed in the ferroaxial susceptibility and the TM shown in Fig.~\ref{f:ratio_dep_Gy}(b) and Fig.~\ref{f:ratio_dep_sx}(b).
These results indicate that the leg inclination serves as a direct geometrical control parameter of the electric field gradient, and hence provides an efficient way to enhance both the ferroaxial moment and the TM.

\section{Summary} 
\label{sec:summary}
We have theoretically investigated the generation of antisymmetric linear TM and ferroaxial moments induced by electric field gradients arising from the geometry of finite systems. 
Considering a tight-binding model on a trapezoidal geometry, we introduced the spatially varying electric field that acts as the conjugate field to the out-of-plane ferroaxial moment.
Based on the Kubo formalism, we showed that the ferroaxial moment is linearly induced by the electric field gradient and exhibits an enhancement at the resonant frequencies. 
We have further clarified that the antisymmetric TM linear in the magnetic field is generated, which is an unconventional response forbidden by Onsager reciprocity in the absence of electric field gradients.
In addition, we showed that the TM is found to be linear in the electric field, whereas the longitudinal magnetization exhibits a quadratic dependence.
This contrast provides a practical advantage, enabling efficient enhancement of the transverse response by tuning the electric field. 
Moreover, we found that both ferroaxial moment and the TM are governed by a common geometric scaling controlled by the leg inclination of the trapezoidal system, which determines the magnitude of the electric field gradient.
This common scaling highlights the electric field gradient as the underlying microscopic origin of both responses.

An important advantage of the present mechanism is its generality: since the electric field gradient is engineered geometrically, it can be realized in a wide class of systems irrespective of crystallographic symmetry.
Our results establish geometry-induced electric field gradients as a versatile control parameter for ferroaxial and transverse responses, enabling the design of unconventional functionalities in mesoscopic and nanoscale systems.

\begin{acknowledgments}
  This research was supported by JSPS KAKENHI Grants Numbers JP22H00101, JP23H04869, and by JST CREST (JPMJCR23O4) and JST FOREST (JPMJFR2366).
  A.I. is financially supported as a JSPS Research Fellow. 
\end{acknowledgments}

\appendix

\section{Ferroaxial moment and transverse magnetization including contribution of uniform electric field}
\label{appendix:w_uniform_E}

\begin{figure}[t!]
  \centering
  \includegraphics[width=\linewidth]{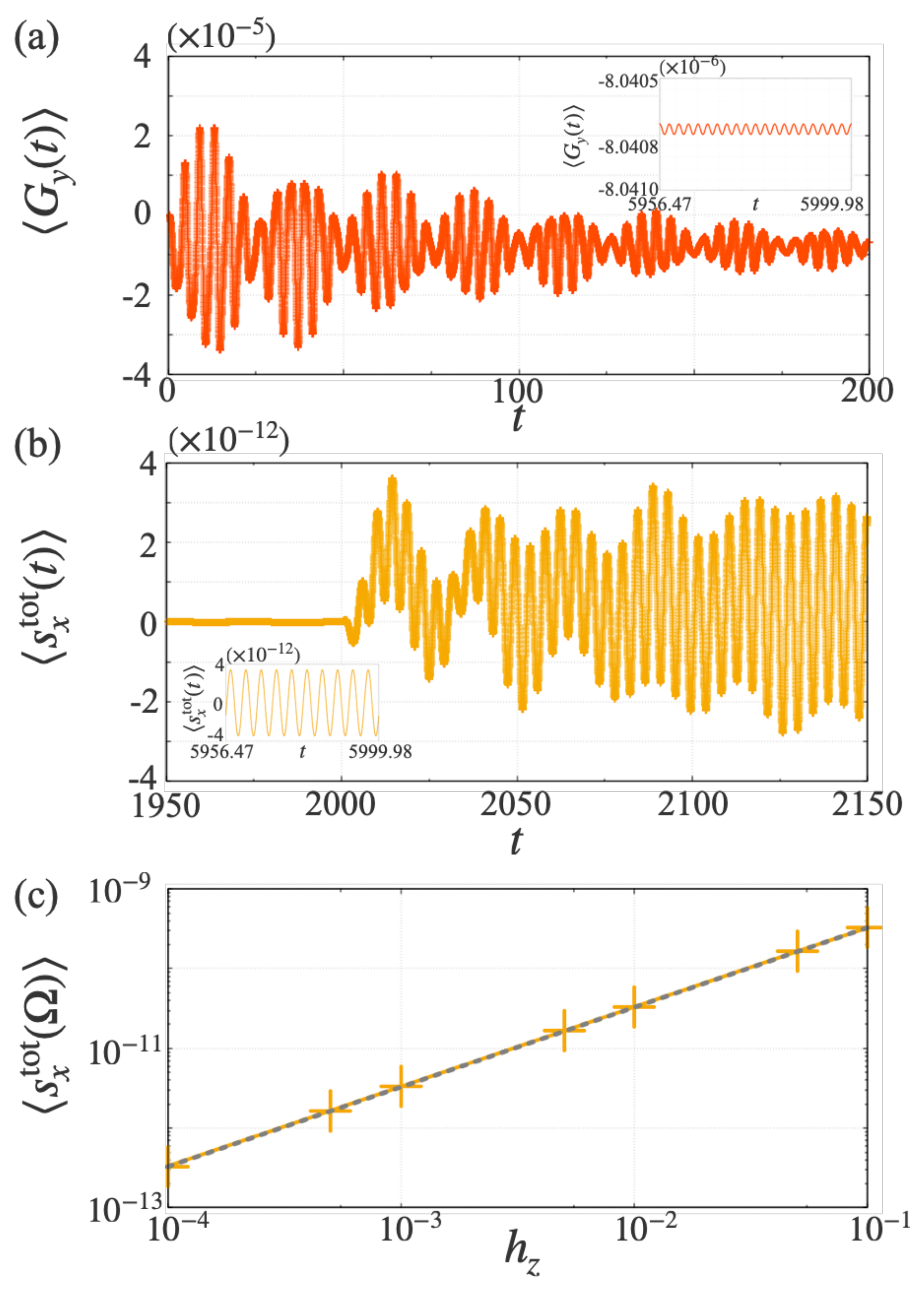}
  \caption{
  \label{f:w_uniform_E_Gy_sx_tdep}
  Same plots as in Fig.~\ref{f:Gy_sx_tdep}, but in the presence of a uniform DC electric field.
  Time evolution of (a) the ferroaxial moment $\langle G_y (t)\rangle$ and (b) the transverse magnetization $\langle s_x^{\rm tot}(t) \rangle$.
  The insets show enlarged views in the interval $5956.47 \leq t \leq 5999.98$, respectively.
  (c) Magnetic field dependence of the transverse magnetization at $\omega=\Omega$.
  The gray dashed line shows a fit to $ax$ with $a=3.27113\times 10^{-9}$.
  }
\end{figure} 
\begin{figure}[t!]
  \centering
  \includegraphics[width=\linewidth]{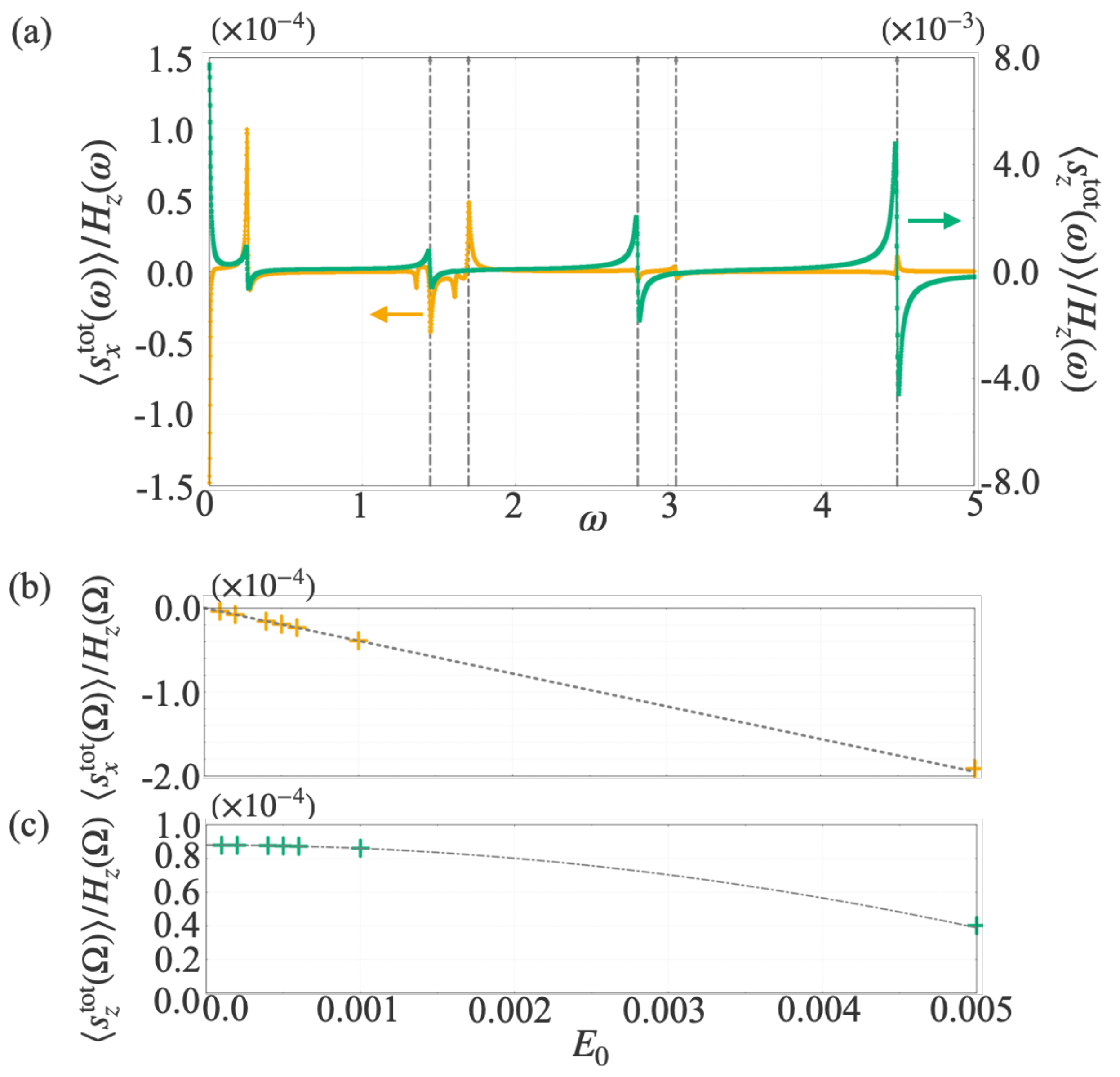}
  \caption{
  \label{f:w_uniform_E_total_tra_longi_mag}
  Same plots as in Fig.~\ref{f:tra_longi_mag}, but in the presence of a uniform DC electric field.
  (a) Frequency dependence of the transverse and longitudinal magnetic responses induced by a pulsed magnetic field along $z$ axis.
  Gray dashed lines indicate the resonant frequencies.
  (b,c) Electric field dependence of the transverse and longitudinal magnetic response at $\Omega=1.44387$, respectively.
  The gray dashed lines show a fit to (b) $ax^b$ with $a=-0.0386845$ and $b=0.998698$ and (c) $ax^b + c$ with $a=-1.94913$, $b=1.99834$, and $c=8.80206\times 10^{-5}$.
  }
\end{figure}

In the main text, we focus on the ferroaxial moment and the TM induced by the electric field gradient, neglecting the contribution from the uniform electric field. 
We show below the response including the contribution from the uniform electric field that is expected to be experimentally observed when an electric field is applied to a trapezoidal sample.
The parameters are set to the same values as those used in Fig.~\ref{f:Gy_sx_tdep} and Fig.~\ref{f:tra_longi_mag} in the main text.

Figures~\ref{f:w_uniform_E_Gy_sx_tdep}(a) and (b) show the time evolution of $G_y$ and the TM under a DC electric field and an oscillating magnetic field, respectively. 
For both the ferroaxial moment and the spin magnetization, though the uniform-field contribution with comparable magnitudes leads to quantitative differences from Figs.~\ref{f:Gy_sx_tdep} (a) and (b), the qualitative behavior remains essentially unchanged: the ferroaxial moment is generated simultaneously with the application of the electric field, while the TM is induced only when the magnetic field is applied in the presence of the electric field. 
In addition, we confirm that the TM is linearly proportional to the electric field as shown in Fig.~\ref{f:w_uniform_E_Gy_sx_tdep}(c).
This result is consistent with the symmetry analysis in the main text; the contribution from the uniform electric field is also linear in the magnetic field.

We also evaluate the contribution from the uniform electric field under an AC electric field. 
Figure~\ref{f:w_uniform_E_total_tra_longi_mag}(a) shows the frequency dependence of the transverse magnetic response $\langle s^{\rm tot}_x(\omega)\rangle  /H_z(\omega)$ and the longitudinal magnetic response $\langle s^{\rm tot}_z(\omega)\rangle /H_z(\omega)$ including the contribution from the uniform electric field.
A qualitatively similar behavior is found for the response to an AC electric field: the uniform-field contributions are comparable to that originating from the electric field gradient, and as shown in Figs.~\ref{f:w_uniform_E_total_tra_longi_mag}(b) and (c), the electric field dependence of the TM and the longitudinal magnetization is linear and quadratic, respectively, as in Figs.~\ref{f:tra_longi_mag}(b) and (c).

\nocite{apsrev42Control}
\bibliographystyle{apsrev4-2}
\bibliography{main}
\end{document}